\documentclass{article}
\usepackage{spconf,amsmath,graphicx}

\usepackage[noadjust]{cite}
\usepackage{booktabs}
\usepackage{multirow}
\usepackage{hyperref} 

\title{A NEURAL TEXT-TO-SPEECH MODEL\\UTILIZING BROADCAST DATA MIXED WITH BACKGROUND MUSIC}
\name{Hanbin Bae, Jae-Sung Bae, Young-Sun Joo, Young-Ik Kim, and Hoon-Young Cho}
\address{Speech AI Lab, NCSOFT Corp., Republic of Korea}
\begin{document}
%
\maketitle
\begin{abstract}
Recently, it has become easier to obtain speech data from various media such as the internet or YouTube, but directly utilizing them to train a neural text-to-speech (TTS) model is difficult. The proportion of clean speech is insufficient and the remainder includes background music. Even with the global style token (GST).
Therefore, we propose the following method to successfully train an end-to-end TTS model with limited broadcast data. First, the background music is removed from the speech by introducing a music filter.
Second, the GST-TTS model with an auxiliary quality classifier is trained with the filtered speech and a small amount of clean speech.
In particular, the quality classifier makes the embedding vector of the GST layer focus on representing the speech quality (filtered or clean) of the input speech.
The experimental results verified that the proposed method synthesized much more high-quality speech than conventional methods.
\end{abstract}
\section{Introduction}
\label{sec:intro}
In recent years, advanced neural text-to-speech (TTS) models have been widely researched \cite{tacotron2, dctts, voiceloop, deepvoice3, transformertts}. 
These are highly dependent on a large speech database, which is not easy to obtain in practice. Recently, as the popularity of personal broadcasting on social networks increases, the opportunity to obtain data on the internet is increasing. However, the amount of clean speech in broadcasted data is typically limited and most part is often noisy due to background music (BGM). Consequently, the importance of constructing a TTS system utilizing noisy speech data is increasing. 

One approach to solve this issue is to preprocess speech data before using it for training the TTS model. \cite{valentini2016speech} and \cite{adiga2019speech} processed noisy speech data using speech enhancement techniques for noise-robust TTS systems.
Speech enhancement research has mainly focused on removing other types of noise rather than BGM, but the primary removal target for broadcasted data is BGM.
In our work, we define a BGM signal as a \textit{music noise}. Methods to remove music noise, including waveform-based or spectral masking-based methods, have been proposed \cite{waveunetss, dcnnss}. However, there is an issue in directly using pre-processed speech and clean speech data for TTS training because the sound quality is slightly different.

Another approach uses a latent variable that embeds the quality of speech. 
The global style token (GST) defines a finite number of tokens and outputs a style embedding vector representing the style of given the reference speech in an unsupervised manner \cite{GST}. In \cite{GST}, it was reported that GST-TTS can be trained utilizing clean, and noisy speech data. GST learns to represent the speech quality condition (clean or noisy) as the reference speech style. 
During the inference step, the GST-TTS system generates a synthesized speech of clear sound by selecting a {\it clean} token.
However, we found experimentally that GST-TTS does not generate synthesized speech successfully when the total amount of speech data is insufficient. Probably because the limited number of tokens makes it difficult to express various types of music and there is no guarantee that one of the tokens can represent {\it clean} speech when the amount of clean speech data is typically limited.

\begin{figure*}[htb]
    \centering
    \centerline{\includegraphics[width=18cm]{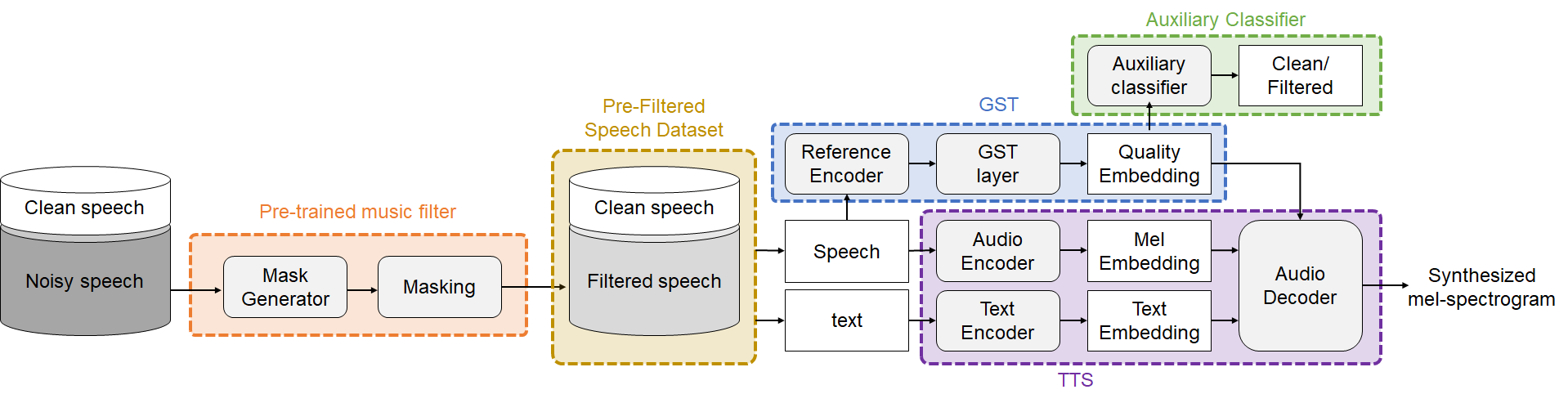}}
\caption{Architecture of the proposed GST-TTS learning method for utilizing a personal broadcast data (training phase).}
\label{fig:proposed}
\end{figure*}

To solve this problem, we propose the GST-TTS learning method to utilize a personal broadcast data.
First, the music noise is removed by introducing a {\it music filter} inspired by \cite{voicefilter}. Next, GST-TTS with an auxiliary quality classifier (AQC) is trained using clean speech and filtered speech (music-removed) from the broadcasted data. GST-TTS can prevent the degradation of synthesized speech caused by using the two types of data together. When the amount of training data is insufficient, the AQC helps the style embedding vector to focus on representing the sound quality of input reference speech rather than the elements of speech such as prosody or speed (we call this the quality embedding vector in the remainder of this paper).
In the inference step, the GST-TTS generates a synthesized speech of clear sound by selecting a clean speech sample as the reference speech.

\section{Related Works}
A similar method, that trains multi-speaker TTS model using crowd-sourced data, was proposed in \cite{hsu2019disentangling}.
Since same speaker's speech is usually recorded in identical environment, they decorrelated speaker identity and noise by using noise-augmented speech data and auxiliary classifiers.
In our proposed method, a music noise is removed first and then an auxiliary quality classifier is adopted for the pre-processed speech and clean speech in TTS training step.

In \cite{voicefilter}, VoiceFilter was proposed to separate target speaker's speech from multi-speakers' speech. 
We introduce a \textit{music filter} inspired by this. The problem is similar in that target speaker's voice is separated from noisy speech. The difference is that in personal broadcast data, there is no need for networks or a feature vector to identify the target speaker because there is no inference audio from other speakers but only music noise.

\section{Proposed Method}
Fig.~\ref{fig:proposed} depicts the proposed learning method. The trained music filter converts the noisy speech (i.e., speech mixed with BGM) data into filtered speech data, after which the GST-TTS with the AQC is trained using the filtered and clean speech data together.

\subsection{Pre-Processing Using the Music Filter}
The left side of Fig.~\ref{fig:proposed} shows the music filtering process.
In the music-filter training phase, the noisy speech dataset is artificially generated by mixing clean speech with randomly selected BGM in a pre-defined signal-to-noise ratio (SNR) range that can be changed according to the target situation. The magnitude spectrogram of the noisy speech is input into the music-filter network and the spectral mask is predicted. The mask is multiplied element-wise to the magnitude spectrogram of the noisy speech to filter out the music noise. 
Then, whole music-filter network is repeatedly updated to reduce the mean squared error between the magnitude spectrograms of the music-filtered-out speech and the original clean speech.
In the inference phase, the target speaker's noisy speech dataset for TTS training is pre-processed using the trained music-filter, i.e., the music noise is filtered out.

\subsection{GST-TTS with the AQC}
The right side of Fig.~\ref{fig:proposed}. depicts the GST-TTS framework with the AQC.
We used deep convolutional TTS (DCTTS) stably trained even under a quite small amount of speech data \cite{dctts}. For the GST layer, we used the same architecture as in \cite{GST}.

In the TTS training phase, the text sequence and corresponding speech having a mel-spectrogram format are input to the TTS model (we refer to a temporally downsampled mel-spectrogram in DCTTS as a mel-spectrogram for general explanation).
The text encoder and audio encoder output the text and audio embedding, respectively, after which an attention matrix is calculated between them.
The mel-spectrogram of input speech is also used as the reference speech to represent the speech quality (clean or filtered). The reference encoder outputs a reference embedding and the GST layer calculates the weights between it and the multiple tokens via a multi-head attention module (we refer to the convex combination of tokens as quality embedding). The quality embedding is concatenated to the context embedding (the product of the attention matrix and text embedding) and is also used as input for the AQC. Finally, the audio decoder estimates the mel-spectrogram by taking the context, quality, and audio embeddings.

To train GST-TTS, the summation of L1 loss ($L_1$) and binary divergence loss ($D_{bd}$) between the estimated and input mel-spectrograms are used \cite{dctts}:
\begin{equation}
    \mathcal{L}_{TTS} = L_{1} + \mathcal{D}_{bd}.
\end{equation}
The AQC includes fully connected networks with one 256 hidden layer and a ReLU activation function \cite{relu} followed by a softmax layer to predict the speech quality. It is trained using binary cross-entropy loss to determine whether the reference speech is clean or filtered.
The final loss function of the proposed learning method can be expressed as 
\begin{equation}
    \mathcal{L}_{total} = \mathcal{L}_{TTS} + \lambda \mathcal{L}_{Aux.}
\end{equation}
where $\mathcal{L}_{Aux.}$ and $\lambda$ are the loss function and loss weight of the AQC, respectively.

In the TTS inference phase, the GST-TTS generates the mel-spectrogram by inputting text and clean speech as the reference speech. Then, the spectrogram super-resolution network (SSRN) of DCTTS predicts a spectrogram from the generated mel-spectrogram and the Griffin-Lim algorithm \cite{glim} estimates phase information. Finally, the time-domain signal is converted from the magnitude and phase of the spectrogram.

\section{Experiments with the Music Filter}
\label{exp_music_filter}

\subsection{Training Setup of the Music Filter}
We used the KsponSpeech \cite{ksponspeech} dataset comprising approximately 1,000 h of spontaneous speech samples recorded with 2,000 speakers to train the speaker-independent music filter. In addition, we collected 68 license-free music samples often used for personal broadcasting from YouTube Studio \cite{youtubestudio}. The music samples were mixed with clean speech with an SNR in the range 0–20 dB. Spectrograms were calculated using a short-time Fourier transform with a window length of 64 ms and a frame interval of 16 ms.

We followed the architectures and hyperparameters of the music filter in \cite{voicefilter} except that batch normalization \cite{batchnorm} was applied to all CNN layers and a ReLU activation function was applied to all layers except for the last one. The networks were trained using the ADAM optimizer \cite{adam} with parameters $lr=0.001$, $({\beta}_1,{\beta}_2)=(0.9, 0.999)$, and ${\epsilon}=10^{-8}$.

\subsection{Performance Evaluation of the Music Filter}
The test set for the music filter comprised 550 speech samples from 11 unseen speakers in a quiet office using mobile devices.)
For the filtered speech dataset, we measured the perceptual evaluation of speech quality (PESQ) \cite{recommendation2001perceptual} and syllable error rate (SER) for pronunciation accuracy. We used a speech recognizer that had 3.48\% SER for the clean speech dataset.

Table~\ref{result:VF-PESQ} indicates that the filtered speech had a higher PESQ score than the noisy speech at each SNR.
Indeed, the filtered speech had a PESQ score of over 3 points at low SNR, which means that the music noise was removed sufficiently well.
The results in Table~\ref{result:VF-SER} of the SER for the clean and filtered speech show that the filtered speech had lower SERs than noisy speech for SNR of 0–5 dB but not for 10–20 dB. At high SNR, distortion caused by the music filter had a more effect on the speech recognizer performance than the relatively low sound noise.

\begin{figure}[t]
    \centering
    \centerline{\includegraphics[width=8.5cm]{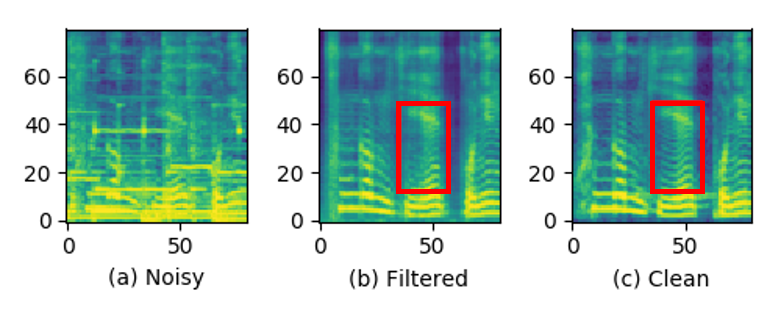}}
\caption{Mel-spectrogram of (a) noisy, (b) filtered, and (c) clean speech samples.}
\label{fig:vf_mel}
\end{figure}

Fig.~\ref{fig:vf_mel} depicts the mel-spectrogram of noisy, filtered, and clean speech samples. Comparing the areas in the red box, it can be observed that the filtered speech was slightly blurred compared to the clean speech, but music noise was clearly removed when compared to the noisy speech.
\begin{table}[t]
  \caption{PESQ scores for noisy and filtered speech}
  \label{result:VF-PESQ}
  \centering
  \begin{tabular}{lrrrrr}
  \toprule
  \multicolumn{1}{c}{\textbf{SNR}} &
  \multicolumn{1}{c}{\textbf{$\mathbf{0dB}$}} &
  \multicolumn{1}{c}{\textbf{$\mathbf{5dB}$}} &
  \multicolumn{1}{c}{\textbf{$\mathbf{10dB}$}} &
  \multicolumn{1}{c}{\textbf{$\mathbf{15dB}$}} &
  \multicolumn{1}{c}{\textbf{$\mathbf{20dB}$}} \\
  \midrule
  Noisy & $2.34$ & $2.63$ & $2.93$ & $3.26$ & $3.59$ \\
  Filtered & $\mathbf{3.12}$ & $\mathbf{3.37}$ & $\mathbf{3.59}$ & $\mathbf{3.78}$ & $\mathbf{3.93}$ \\
  \bottomrule
  \end{tabular}
\end{table}
\begin{table}[t]
  \caption{SER (\%) for noisy and filtered speech}
  \label{result:VF-SER}
  \centering
  \begin{tabular}{lrrrrr}
  \toprule
  \multicolumn{1}{c}{\textbf{SNR}} &
  \multicolumn{1}{c}{\textbf{$\mathbf{0dB}$}} &
  \multicolumn{1}{c}{\textbf{$\mathbf{5dB}$}} &
  \multicolumn{1}{c}{\textbf{$\mathbf{10dB}$}} &
  \multicolumn{1}{c}{\textbf{$\mathbf{15dB}$}} &
  \multicolumn{1}{c}{\textbf{$\mathbf{20dB}$}} \\
  \midrule
  Noisy & $51.57$ & $22.88$ & $\mathbf{8.91}$ & $\mathbf{4.95}$ & $\mathbf{3.65}$ \\
  Filtered & $\mathbf{24.23}$ & $\mathbf{14.70}$ & $9.21$ & $7.61$ & $6.79$ \\
    \bottomrule
  \end{tabular}
\end{table}
%
%
%
%

\section{Performance Evaluation}
Ablation tests were conducted to validate the effect of the proposed method. We compared the following five models:
\begin{itemize}
  \setlength\itemsep{0em}
  \item \textit{TTS}: the TTS model (DCTTS)
  \item \textit{GST}: the TTS model with quality embedding
  \item \textit{GST+Aux.}: \textit{GST} with the AQC
  \item \textit{GST+MF}: \textit{GST} with the music filter
  \item \textit{GST+MF+Aux.}: \textit{GST+MF} with the AQC
\end{itemize}
\textit{GST} and \textit{GST+Aux.} were trained using the clean and \textit{music-mixed} speech dataset while \textit{GST+MF} and \textit{GST+MF+Aux.} were trained using the clean and \textit{music-filtered} speech dataset.

\subsection{Training Setup}
We trained the models for various ratios of clean and noisy/filtered speech data. We used approximately 5 h of speech data recorded from a single Korean female speaker at 16 kHz sampling frequency and artificially generated noisy and filtered speech datasets using the clean speech data as described in the previous section. After excluding 1\% of the 5 h speech dataset as the test set, we configured the training set so that the amount of clean speech was $0.5$, $1.5$, and $2.5$ h and the remainder was noisy or filtered speech.
Text input is a character sequence and input speech and reference speech are 80-bin mel-spectrograms computed with a fast Fourier transform size of 1024, a hop size of 256, and a window size of 1024.

For the TTS network, we followed the same network architecture and hyperparameters as in \cite{dctts}. except that we trained the SSRN as a universal model using a multi-speaker speech corpus recorded by 66 speakers (1 h for each speaker) because the amount of clean speech was insufficient (e.g., 0.5 h) to train the SSRN.
For the GST network, we used the same architecture (except for the TTS model) and hyperparameters as that in \cite{GST}. The number of style tokens and heads of multi-head attention were set to 10 and 4, respectively, as was carried out in \cite{GST}.
The loss weight of the AQC, $\lambda$, was $0.001$, $0.01$, and $0.01$ for $0.5$, $1.5$, and $2.5$ h amounts of clean speech, respectively.
We set a lower loss weight value for $0.5$ h of clean speech so that the AQC learned slowly. For $\lambda=0.01$, the AQC converged quickly because the data imbalance problem was worse than with the other two cases. And the quality embedding did not affect the synthesized sample.

\begin{figure}[t]
    \centering
    \centerline{\includegraphics[width=8.5cm]{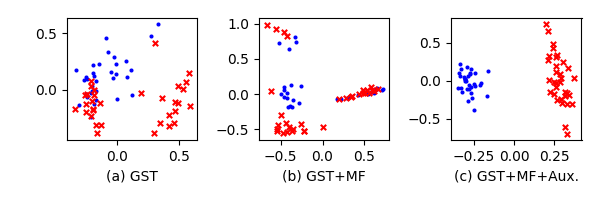}}
\caption{PCA plots of the quality embedding. The blue dots denote clean speech and the red ‘X’s denote noisy or filtered speech.} 
\label{fig:pca}
\end{figure}
\subsection{Visualization of the Quality Embedding }
We visualized the quality embedding using principal component analysis (PCA) \cite{hotelling1933pca}. The quality embedding was extracted from models trained with 0.5 h clean speech.
As shown in Fig~\ref{fig:pca}, only the \textit{GST+MF+Aux.} model clustered the clean and filtered speech clearly.
It means that the AQC is crucial for ensuring that the quality embedding represents the quality of reference speech by separating the clean and filtered speech.

\subsection{Pronunciation Accuracy}
\label{ssec:eval}
We measured the SER to confirm the intelligibility of the synthesized speech. 
Unfortunately, \textit{GST} and \textit{GST+Aux.} using clean and noisy speech were not trained successfully because the training loss diverged. Accordingly, \textit{TTS}, \textit{GST+MF}, and \textit{GST+MF+Aux.} were evaluated. As the baseline, SER values for \textit{TTS} trained with 5 h clean speech and filtered speech were $13.45$\% and $74.12$\%, respectively.
The other results are listed in Table~\ref{result:GST-SER}.
Both of the other models showed a tendency for SER to decrease as the amount of clean speech data increased. The results for \textit{GST+MF} were extremely poor in all cases, while even when trained with $1.5$ h of clean speech, the SER for \textit{GST+MF+Aux.} was only $2.44$\% higher than for \textit{TTS} trained with $5$ h of clean speech.

\begin{table}[t]
  \caption{SER (\%) results for synthesized speech.}
  \label{result:GST-SER}
  \centering
  \begin{tabular}{lccc}
  \toprule
  \multicolumn{1}{c}{\textbf{Clean / Filtered (h)}} &
  \multicolumn{1}{c}{\textbf{$\mathbf{0.5 / 4.5}$}} &
  \multicolumn{1}{c}{\textbf{$\mathbf{1.5 / 3.5}$}} &
  \multicolumn{1}{c}{\textbf{$\mathbf{2.5 / 2.5}$}} \\
  \midrule
GST+MF      & $74.40$ & $58.13$ & $49.12$ \\
GST+MF+Aux. & $\mathbf{29.80}$ & $\mathbf{15.89}$ & $\mathbf{16.52}$ \\
    \bottomrule
  \end{tabular}
\end{table}

\subsection{Subjective Evaluation}
We conducted mean opinion score (MOS) tests on speech quality and naturalness. Note that the models that had a higher SER than $50$\% were not used for the MOS test because the synthesized speech from these models was either very noisy or difficult to understand.
Sixteen native Korean speakers participated and were asked to give scores from $1$ (bad) to $5$ (excellent). Fifty synthesized speech samples (10 samples for each model) were randomly played.

Table \ref{result:GST-MOS} reports the MOS test results. 
The larger the amount of clean speech data, the higher the MOS scores for both quality and naturalness. Especially, the proposed \textit{GST+MF+Aux.} model trained with $2.5$ h of clean speech achieved the highest scores of $3.92$ and $3.85$, respectively, which were close to those of \textit{TTS} trained with $5$ h of clean speech. Moreover, \textit{GST+MF+Aux.} trained with $0.5$ h of clean speech outperformed \textit{GST+MF} trained with even $2.5$ h of clean speech.

Audio samples can be found online\footnote{https://nc-ai.github.io/speech/publications/tts-with-bgm-data/}.

\begin{table}[t]
  \caption{MOS test results for speech quality and naturalness with 95\% confidence intervals. `C' and `F' denote the amount of clean and filtered speech, respectively.}
  \label{result:GST-MOS}
  \centering
  \begin{tabular}{lrcc}
  \toprule
  \multicolumn{1}{c}{\textbf{Model}} &
  \multicolumn{1}{c}{\textbf{C/F (h)}} &
  \multicolumn{1}{c}{\textbf{Quality}} & 
  \multicolumn{1}{c}{\textbf{Natural}} \\
  \midrule 
    TTS             & $5$ / 0   & $4.04 \pm 0.12  $ & $4.20 \pm 0.13 $ \\ 
  \midrule
    GST + MF        & $2.5$ / 2.5  & $1.84 \pm 0.10$ & $2.06 \pm 0.15$ \\
  \midrule  
    \multirow{3}{*}{\shortstack[l]{GST + MF \\ \hspace{0.66cm} + Aux.\\(Proposed)} }& $0.5$ / 4.5& $3.05 \pm 0.14$ & $2.91 \pm 0.16$ \\
    & $1.5$ / 3.5 & $3.87 \pm 0.13$ & $3.74 \pm 0.15$ \\
    & $2.5$ / 2.5 & $\mathbf{3.92 \pm 0.12}$ & $\mathbf{3.85 \pm 0.14}$\\

    \bottomrule
  \end{tabular}
\end{table}

\section{Conclusions}

We proposed a learning method of TTS that can generate clean synthesized speech under the limitation of personal broadcast data. To successfully train the TTS model, the music noise is removed by introducing a music filter and GST-TTS with the AQC is trained using the filtered speech and a small amount of clean speech. The AQC makes the quality embedding be effectively learned to represent the speech quality. The model learned by the proposed method generated natural and intelligible speech using a small amount of clean speech data that was almost comparable to baseline TTS trained using much more speech data.

\bibliographystyle{IEEEbib}
\bibliography{refs}
\end{document}